# Fall Rate of Sub-micron Particles and Bare Shear Viscosity: I. Diffusion Equation

**R. L. Varley**

Hunter College of the City University of NY

Department of Physics and Astronomy

695 Park Avenue, New York, NY 10065 USA

Rodney.Varley@Hunter.Cuny.Edu

Kim and Fedele discovered experimental evidence for the breakdown of the Millikan's Law for the fall rate of oil droplets in Nitrogen gas and the discrepancy is most pronounced for smallest, sub-micron size particles. Here we explain these results by showing that the particle's motion is determined in part by the bare shear viscosity $\eta_\lambda$ which is defined by the averaging length $\lambda$. This is in contrast with the usual theory which involves the renormalized shear viscosity $\eta$. An increase in gas pressure produces a decrease in the bare shear viscosity and as a result, the fall rate increases. This behavior is opposite the Millikan law prediction that an increase in pressure produces a decrease in fall rate. As a result, the bare shear viscosity $\eta_\lambda$ is experimentally measurable by the fall rate. The theory here uses a convective diffusion equation and a Langevin approach will be presented elsewhere.

51.20.+d, 05.40.+j, 05.20.Dd, 51.10.+y

## 1. Introduction

The fall rate of particles of sub-micron size is of interest in a number of scientific areas [1]. Particles of this size are comparable to the wavelength of light and the interaction between light and these particles is especially strong. Furthermore, since particles of this size also tend to settle out of the atmosphere at a much slower rate than larger particles, their effect on sunlight is much longer lived. As an example, there is an indication that an impact of large meteors with the Earth could generate large quantities of small particles which would shield the Earth from sunlight and it is thought this may have caused the final demise of the dinosaurs [2]. More recently, the smoke generated by the uncontrolled burning of oil in Kuwait after Desert Storm introduced large quantities of sub-micron particles high into the atmosphere. Particles of the sub-micron size also have been implicated in causing lung cancer since among other things, the lungs have difficulty discharging very small particles. A great deal of effort has gone into the development of electrostatic precipitators etc. so that these particles are removed from the smoke of coal fired power plants.

Many of the predictions of the rate particles settle out of the atmosphere are based on the empirical Millikan's law [3]. This law was originally developed in connection with the famous Millikan oil drop experiment for the determination of the charge on an electron. The particles studied experimentally by Millikan were of a radius larger than 0.245 $\mu$M and gas pressure at most one atm and often quite less. More recently, Kim and Fedele [4] studied the fall rates of smaller particle down to a radius of perhaps 0.1 $\mu$M and pressures from 1-15 atm. For the larger sized particles, Kim and Fedele found the fall rate decreased when the gas pressure increased in agreement with Millikan and the idea that the boundary conditions change from slip to stick. However, for the smaller particles they found the fall rate *increased* when the gas pressure is increased. The Kim-Fedele experiments were done with up most care. The protocol of using the *same drop* to measure the fall rate at various pressures and the reproducibility of the results make them especially convincing. The oil drops studied by Kim and Fedele are so small that Brownian motion becomes obviously apparent and given the anomalous behavior of the fall rate, it is surprising that the mean square displacement measured by Kim and Fedele behaves normally as a function of gas pressure [4].

Here we outline a fall rate theory to see what ingredients can lead to an explanation of the Kim-Fedele experiments. This new fall rate theory is fundamentally based on the idea that the shear viscosity $\eta_\lambda$ for a gas depends upon the size or scale of the hydrodynamic phenomena of interest provided the scale is small enough [5]. Section 2 introduces the bare shear viscosity $\eta_\lambda$ as related to the Navier-Stokes equation together with the concepts of averaging length $\lambda$ and measurement length $\Lambda$. The renormalized shear viscosity $\eta$ is what is usually measured in the laboratory and $\eta$ applies to usual large scale hydrodynamic experiments. We propose that the bare shear viscosity $\eta_\lambda$ applies to small scale fluid flow which is being probed by the sub-micron oil drops in the Kim and Fedele experiment. Here the connection between $\eta_\lambda$ and $\eta$ is given by renormalization theory and an interpolation function obtained using a mode coupling calculation is mentioned. Section 3 gives the intuitive rationale for involving the bare shear viscosity $\eta_\lambda$ in the drag force on small spheres [6] falling in a gas. Our argument is based on the contemplation of a numerical solution of the flow together with the calculation of the drag force using the stress tensor. We do not pretend our argument is rigorous for a Stokes-like drag force which has the bare shear viscosity $\eta_\lambda$ instead of the usual $\eta$. However, it should also be kept in mind the Millikan formula as a correction to Stoke's Law has an empirical basis so it is entirely appropriate that the discussion in section 3 is not rigorous. A convective diffusion equation for the Brownian or B-particle concentration in an external field is presented in Section 4 and this depends upon the bare diffusion coefficient and bare mobility in keeping with accepted practice. The results of a fluctuation-renormalization calculation produces a renormalized diffusion coefficient however the bare mobility is not affected. The fall rate is obtained from the convective diffusion equation in section 5 and a comparison is made with both the Kim-Fedele experimental results as well as the Millikan theory. Section 6 shows how to calculate the mean square displacement of the B-particle and that the theory agrees with the Millikan result in this respect. Finally our results are summarized in Section 7 there also is a discussion of possible future work.

## 2. The Navier-Stokes Equation and Renormalization

The fluid velocity $V(r, t)$ at the spatial point $\mathbf{r}$ and time t is usually defined in hydrodynamics [5, 6, 7, 8] as the average of the velocities of the gas molecules contained in a small imaginary volume $\lambda^3$ which is a part of the total system and it is reasonable to call $\lambda$ the "averaging length" [5]. The volume $\lambda^3$ for the computation of $V_\lambda(r, t)$ is usually taken large enough so that $n_0 \lambda^3$ contains a large number of gas molecules and the fluctuation in the number of particles in $n_0 \lambda^3$ is small. If the mean distance $\ell$ between the molecules of the fluid is obtained from the equilibrium number density $n_0 = (1/\ell^3)$ then the condition that $n_0 \lambda^3$ contains a large number of molecules means $\ell \ll \lambda$. Some authors [8] include an additional average over a time long compared with the collision time especially when $n_0 \lambda^3$ contains relatively few particles and it is assumed the value of the fluid velocity $V_\lambda(r, t)$ is independent of the size and shape of the volume $\lambda^3$ which is reasonable provided $\lambda$ is large and for this reason the fluid velocity will be referred to simply as $V(r, t)$ but the $\lambda$ is implicitly there. It is important to keep in mind that $\lambda$ cannot be taken too large either without missing important phenomena due to the averaging but aside from these conditions the precise numerical value of $\lambda$ is usually not specified.

There is a second, larger length scale $\Lambda$ which is used in principle to calculate the actual behavior of the fluid for example, as it flows passed a sphere. $\Lambda$ can be no smaller than $\lambda$ because after doing the averaging mentioned above, there is no variation of the fluid velocity on a scale smaller than $\lambda$. When doing usual hydrodynamic calculations quite often the size of $\Lambda$ can be ignored because the results are insensitive to the size of $\Lambda$. The length $\Lambda$ we have in mind is much smaller than the B-particle radius a that is, $\Lambda \ll a$ since $\Lambda$ is a length scale that is small enough to capture the *significant* variation in the fluid velocity. Many hydrodynamic systems studied in the past were large enough so that the conditions $\ell \ll \lambda$ and $\Lambda \ll a$ together with $\lambda \leq \Lambda$ were satisfied easily but for sub-micron drops there can be a problem and an accommodation must be made if we are to use a hydrodynamic description. Incidentally there is a whole class of problems of this sort involving MEMS or Micro-Electro-Mechanical-Systems [9] which are of considerable current interest. Later in section 3, we will make a pragmatic choice $\lambda = \Lambda$ which is consistent with the conditions on $\lambda$ and $\Lambda$. However, for now it is useful to keep the conceptual difference between $\lambda$ and $\Lambda$.

The drag force on the Brownian particle given here assumes the surrounding fluid is incompressible so $\mathbf{V}(\mathbf{r}, t)$ satisfies the Navier-Stokes equation including $\rho g$ the external force of gravity [6]

$$\rho \frac{\partial}{\partial t} \mathbf{V}(r, t) + \rho (\mathbf{V}(r, t) \cdot \nabla) \mathbf{V}(r, t) = -\nabla p(r, t) + \eta_\lambda \nabla^2 \mathbf{V}(r, t) + \rho \mathbf{g} \tag{1}$$

$V(r, t)$ is the fluid velocity, the pressure in the fluid is $p(r, t)$ while $\rho = m\, n_0$ is the fluid mass density where m is the mass of a molecule of the gas and g is the acceleration of gravity. The so-called "bare" shear viscosity $\eta_\lambda$ for the averaging length $\lambda$ is introduced here consistent with modern usage [10, 11, 12]. The Fourier transform of the spatial variable **r** appearing in the fluid velocity $\mathbf{V}(r, t)$ is given by

$$\mathbf{v}_\kappa(t) = \int \mathbf{V}(r, t)\, e^{i\kappa \cdot r}\, dr \tag{2}$$

When the fluctuations in the fluid density, fluid velocity etc. are large because the averaging length $\lambda$ is small, then equation (1) is augmented by a fluctuating stress tensor term that has a zero average [6]. This plays a role similar to the fluctuating force in the Langevin equation. The fluctuations play an important role since a stochastic average of the nonlinear, convective term in equation (1) results in a "fluctuation-renormalization" term $-R_\lambda^\eta(\kappa)\,\kappa^2\,\overline{\mathbf{v}}_\kappa(t)$ and equation (1) becomes after the Fourier transform [10, 11, 12, 13]

$$\rho\,\frac{\partial}{\partial t}\overline{\mathbf{v}}_\kappa(t) = -i\,\kappa\,\overline{p}_\kappa(t) - \eta(\kappa)\,\kappa^2\,\overline{\mathbf{v}}_\kappa(t) \tag{3}$$

$\overline{\mathbf{v}}_\kappa(t)$ in equation (3) is stochastic average fluid velocity for the wavenumber $\kappa$ and also $\overline{\mathbf{v}}_\kappa(t)$ is defined with the steady state solution removed that is due to gravity. As an aside, the steady state solution reappears as the familiar buoyant force $m_d g$ on the B-particle where $m_d$ is the fluid displaced by the B-particle. However, for the Kim-Fedele experiments the B-particle is massive $M \gg m_d$ so the buoyant force is small compared with the gravitational force and so it will neglected. The Fourier transform of the nonlinear convective term for the average velocity $\overline{V}(r, t) \cdot \nabla \overline{V}(r, t)$ was not included in equation (3) because the Reynolds number Re = Ua/$\nu$ is small for sub-micron particles.

The fluctuation-renormalization term is added to the Fourier transform of the viscous term $-\eta_\lambda\,\kappa^2\,\overline{\mathbf{v}}_\kappa(t)$ in equation (1) so that the "renormalized" shear viscosity $\eta(\kappa)$ for the measurement wavenumber $\kappa$ is given by [10, 11, 12, 13]

$$\eta(\kappa) = \eta_\lambda + R_\lambda^\eta(\kappa)\ . \tag{4}$$

in equation (3). This mixing in equation (4) of the measurement wavenumber $\kappa$ and the averaging length $\lambda$ is a little unfortunate but has the virtue of distinguishing the two length $\lambda$ and $\Lambda$ and their associated wavenumbers $k = 2\pi/\lambda$ and $\kappa = 2\pi/\Lambda$. Often the term "renormalized" is reserved for the very large scale shear viscosity $\eta(0)$ where $\Lambda \to \infty$ or $\kappa \to 0$ and by the way, it is easy to show from the definition of $\eta_\lambda$ that $\eta_\lambda \to \eta(0)$ when $\lambda \to \infty$. The "fluctuation-renormalization" $R_\lambda^\eta(\kappa)$ interpolates between the shear viscosity $\eta_\lambda$ for the averaging length $\lambda$ and the renormalized, large scale shear viscosity $\eta(\kappa)$ where $\Lambda = 2\pi/\kappa$. When $\Lambda=\lambda$ (or $\kappa$=k) there is no need for interpolation and $R_\lambda^\eta(\kappa)$ =0 so that $\eta(\kappa) = \eta_\lambda$ so in other words, the bare shear viscosity $\eta_\lambda$ equals the renormalized shear viscosity $\eta(\kappa)$ when it is measured on a length scale $\Lambda$ equal to the averaging length scale $\lambda$. An important special case of equation (4) which we use is when $\kappa$=0 then $\eta(0) = \eta_\lambda + R_\lambda^\eta(0)$. We will take the large scale shear viscosity $\eta(0)$ to equal the Enskog shear viscosity $\eta$E and $\eta(0) \simeq \eta$E is a very good approximation at least for pressures in the range of 1-15 atm used in the Kim-Fedele experiments. Combining the above results in a formula for the bare shear viscosity $\eta_\lambda = \eta$E $- R_\lambda^\eta(0)$ which we will have use for shortly. A straight forward but tedious mode coupling calculation [14] results in $R_\lambda^\eta(0) = \alpha\, k_B\, T\, \rho/\eta_\lambda\, \lambda$ with $k_B$ the Boltzmann constant, T the gas temperature, $\rho$ the gas mass density, and $\alpha = 14/5\,\pi^2$ is a numerical constant. Details of this calculation will be given elsewhere and it should be kept in mind this is an approximate result taking into account the size of the Enskog values of the transport coefficients in the pressure range 1-15 atm. Experience with computer simulations [22] indicates this value underestimates $R_\lambda^\eta(0)$ probably since the ordinary hydrodynamic equations are inaccurate at short times and small distances. But none-the-less this $R_\lambda^\eta(0)$ will be used here to yield a practical equation for $\eta_\lambda$

$$\eta_\lambda = \eta_E - \frac{\alpha\, k_B\, T\, \rho}{\eta_\lambda\, \lambda}\ . \tag{5}$$

provided we make a good choice for the size of $\lambda$ and more will be said about this in the next section. Equation (5) is a quadratic which may be solved for $\eta_\lambda$ or alternatively, equation (5) can be solved iteratively for $\eta_\lambda$ and in the first iteration, $\eta_E$ appears in the denominator on the right hand side.

It is generally expected that $\eta_\lambda \to 0$ as $\lambda \to 0$ because on the microscale the system is not dissipative. Notice there is an averaging length $\lambda$ for which the right hand side of equation (5) vanishes and thus $\eta_\lambda=0$ however, this $\lambda$ is almost certainly too large since it is much larger than the mean molecular distance. Also as $\lambda$ decreases further, $\eta_\lambda$ turns negative (or worse complex if the quadratic form is used) but this is not physically possible. Corresponding statements can be made about the gas density $\rho$ and as a result equation (5) can be applied only for pressures less than 15 atm when particles as small as 0.2 $\mu$M particles are considered. The density range for $\eta_\lambda$ can be extended with a simple Padé approximation

$$\eta_\lambda = \frac{\eta_E}{\left(1 + \frac{\alpha k_B T \rho}{\eta_E^2 \lambda}\right)} . \tag{6}$$

which has the proper behavior $\eta_\lambda \to 0$ as $\lambda \to 0$ but probably at a cost of accuracy at higher pressure. The above method for obtaining the bare shear viscosity $\eta_\lambda$ in equation (5) is indirect since $\eta_\lambda$ is given in terms of the renormalized shear viscosity $\eta(0) \simeq \eta_E$ together with the renormalization $R_\lambda^\eta(0)$. There is another direct way of calculating $\eta_\lambda$ as the time integral of the Peculiar Stress Tensor Auto-Correlation Function or PSTACF in much the same way that the bare diffusion coefficient $D_\lambda$ has been obtained in terms of the time integral of the Peculiar Velocity Auto-Correlation Function or PVACF [5]. This method of obtaining $\eta_\lambda$ is especially useful when combined with computer molecular dynamic simulation but we will not use this method here.

# 3. A Modified Stokes Law

Fall rate calculations require the drag force, or resistance of the fluid on the spherical B-particle. One might use the generalized Faxen theorem for drag force on a sphere in an incompressible fluid with stick boundary conditions for the general case of non-steady motion of the sphere together with a fluid with fluctuations [15]. However, this is quite a bit more complicated than is required here and we note the generalized Faxen theorem reduces to the familiar Stokes law $F = 6\pi \eta a U$ in the case where the B-particle mass density (oil) is large compared with the mass density of the surrounding fluid (Nitrogen) as is the case in the Kim-Fedele experiment. U is the velocity of the spherical B-particle, a is its radius, and it is assumed that the fluid comes to rest with respect at the B-particle surface (the so-called "stick" boundary conditions). The "stick" boundary conditions apply only at gas pressure large enough so that the mean free path $\mu$ is small compared with the B-particle radius a and the "slip" boundary condition correction [1, 16] becomes important when $\mu \approx a$. Millikan [3] proposed a modified Stokes law $F = 6\pi\eta a U/\xi$ where the empirical interpolation function $\xi$ between slip and stick boundary conditions is $\xi = 1 + A(\mu/a) + B(\mu/a) e^{-C(a/\mu)}$. A modern analysis [17] of Millikan's original data provides A=1.155, B=0.471, and C=0.596 for oil droplets in air and the mean free path $\mu$ is calculated using $\mu = \eta/(0.490874 \rho \bar{c})$ where $\bar{c} = \sqrt{(8/\pi) k_B T/m}$ the most probable speed of the gas molecules, $\rho$ is the mass density of the gas, and we use $\eta = 1.656 \times 10^{-4}$ gm/(cm − sec) since the Kim-Fedele experiments were in Nitrogen. It is well known that $\eta$ is independent of pressure [8] over a wide pressure range from $10^{-2}$ atm to a least 20 atm.

Here we suggest that the oil droplets in the Kim-Fedele experiment are so small, that for the purposes of calculating the drag force, some account of the wave number $\kappa$ variation in $\eta(\kappa)$ appearing in the Navier-Stokes equation (3) has to be made. The usual drag force theory consists of taking $\eta(\kappa)$ to be the Enskog or large scale shear viscosity and this is an appropriate approximation for larger spheres. For smaller spheres, we use the following intuitive argument to estimate the size of $\Lambda=2\pi/\kappa$. Suppose we imagine a numerical solution of the Navier-Stokes equation for the fluid velocity on a grid of cubes each having volume $\Lambda^3$ and this is followed by a numerical integration of the stress tensor over the surface of the sphere to get the drag force [6]. By the way, a numerical solution is necessary if the B-particle has an irregular shape unless of course it is approximated by a sphere. Clearly $\Lambda$ must be much smaller than the radius a of the B-particle otherwise important variations of the hydrodynamic field in the calculation of the drag force are averaged away. The numerical error due to the size of $\Lambda$ can be estimated using the Stokes analytical solution in the stress tensor followed by a numerical integration over the surface of the sphere to get the drag force [6]. Symmetry reduces the surface integral to a single $\theta$ angular integration and it quickly becomes clear the angular distance between data points $\Delta\theta$ can be no larger than $(\pi/2)/100 \simeq 0.01$ in order to get the drag force accurate to say 1% with the corresponding $\Lambda \simeq a\Delta\theta \simeq 0.01\, a$ roughly. One might think it a good idea to take $\Lambda$ smaller still since that would seem to improve the accuracy of the drag force calculation but then the fluctuations in the fluid velocity would become so large that the average velocity would not have much meaning. The fluctuations on a small scale are important but their effect appears in the fluctuating force acting on the B-particle while the drag force corresponds to the average behavior. As a result of the above considerations, the Stokes law drag force is approximated by $F = 6\pi\eta(\kappa) aU/\xi$ with $\kappa = 2\pi/\Lambda$ and $\Lambda = 0.01\, a$ provisionally. Since $\ell \ll \lambda \leq \Lambda \ll a$ and we have trouble satisfying both $\ell \ll \lambda$ and $\Lambda \ll a$ for sub-micron particles, a pragmatic choice $\lambda = \Lambda$ and finally this condition simplifes the drag force to

$$F = 6\pi \eta_\lambda\, a\, U/\xi \tag{7}$$

with $\lambda = 0.01\,a$ and where the same $\xi$ is used as was proposed by Millikan to account for the transition from stick to slip boundary conditions. Obviously using Millikan's $\xi$ unchanged is a bit problematic since Kim-Fedele worked in a pressure and B-particle size regime outside that studied by Millikan. That said, our choice of $\eta_\lambda$ to approximate $\eta(\kappa)$ is made in the same empirical spirit Millikan used to construct $\xi$. Incidentally $\lambda=0.01a$ also works fairly well for particles having a larger radius a since then $\lambda=0.01a$ is so large that $\eta_\lambda \to \eta(0)$. The above argument is used only to get an estimate of the size of $\lambda$ but in section 5 an argument is made that it is the experiment itself that determines the size of the parameter $\lambda$ through the experimental determination of the bare shear viscosity $\eta_\lambda$. The bare particle mobility $b_\lambda$ defined by $U = b_\lambda F$ and equation (7) yields

$$b_\lambda = \xi/(6\pi\eta_\lambda\, a) \tag{8}$$

and this mobility will be used in the convective diffusion equation discussed in the next section.

## 4. The Convective Diffusion Equation

The dynamical behavior of the concentration $\mathbf{c}(r,t)$ of the B-particle solute in a Nitrogen gas solvent and acted on by the gravitational force Mg is well established [6, 18, 19, 20] and given by

$$\rho\frac{\partial}{\partial t}\mathbf{c}(r,t) + \rho(\mathbf{V}(r,t)\cdot\nabla)\mathbf{c}(r,t) = -\nabla\cdot\mathbb{J} \tag{9}$$

where the concentration current $\mathbb{J}$ relative the solvent is given by

$$\mathbb{J} = -\rho\, D_\lambda\, \nabla \mathbf{c}(r,t) + \rho\, b_\lambda\, M\, g\, \hat{z}\, \mathbf{c}(r,t) + \mathbb{J}_R \tag{10}$$

$\rho$ is the total mass density (solvent plus solute) and it is accepted practice [18, 20] to call the $D_\lambda$ that appears in the constitutive relation (10) the "bare" diffusion coefficient. The "bare" mobility $b_\lambda$ is the effect of the gravitational acceleration g acting in the $\hat{z}$ direction on the B-particle mass M. The term "bare" mobility $b_\lambda$ appears not to have been used before but it's use here is consistent with $D_\lambda$. $\mathbb{J}_R$ is a random current which accounts for the fluctuations in the fluid [6, 18] when $\lambda$ is relatively small. The combination of equations (9) and (10) results in a convective diffusion equation for the B-particle system in gravity. The B-particle problem involves two times scales, the shorter of which is the "relaxation time" $M\xi/(6\pi\eta a)$ of the B-particle velocity distribution to an equilibrium Maxwell-Boltzmann form. The "diffusion time" scale $\mathcal{L}^2/D$ where D is the diffusion constant is much longer and is physically the time it take the concentration to spread out a spatial distance $\mathcal{L}$ and $\mathcal{L}\gg a$. The larger the spatial region $\mathcal{L}$, the longer the diffusion time. When considering the longer time scale of diffusion, it is appropriate to use the steady state Stokes-Millikan form $b_\lambda=1/(6\pi\eta_\lambda\, a)$ for the "bare" mobility.

A "fluctuation renormalization" [13, 14] of the convective term in equation (9) plus (10) leads to a diffusion equation of the form [18, 20]

$$\frac{\partial}{\partial t}\bar{c}_\kappa(t) = -D[\kappa]\,\kappa^2\,\bar{c}_\kappa(t) + i\,\kappa_z\, b_\lambda\, m_B\, g\, \bar{c}_\kappa(t) \tag{11}$$

where the spatial Fourier transform of the fluctuation average concentration $\bar{c}_\kappa(t)$ is defined similarly to equation (2) with $\kappa_z$ the wavevector component associated with the z coordinate and $D[\kappa]$ is a "renormalized" diffusion coefficient. $D[\kappa]$ is related to the bare diffusion coefficient $D_\lambda$ by the renormalization equation [18, 20] $D[\kappa] = D_\lambda + R_\lambda^D[\kappa]$ where the renormalization $R_\lambda^D[\kappa]$ interpolates between $D_\lambda$ measured on the scale $\lambda=2\pi/k$ and $D[\kappa]$ measured on the scale $\Lambda=2\pi/\kappa$. Notice that while the diffusion coefficient is fluctuation renormalized, the mobility is not renormalized and it is the bare mobility that appears in equation (11). Usually diffusion on a large scale $\Lambda\to\infty$ (or $\kappa\to 0$) is what is usually measured or observed. What this means is that $\Lambda$, the scale on which the concentration varies is much larger than $\lambda$ and in practice $\Lambda\gg a$ the B-particle radius and what we need is the special case when $\kappa\to 0$

$$D[0] = D_\lambda + R_\lambda^D[0] \tag{12}$$

The measurement length $\Lambda$ for diffusion can be different from the measurement length for the fluid velocity (discussed previously) since different measurements are involved. So here we take $\Lambda \to \infty$ (or $\kappa \to 0$) since this limit appears naturally when calculating the mean square displacement and the fall rate as will be seen shortly. Several authors [18, 20] have pointed out $D_\lambda = 0$ for large, non-molecular sized B-particles. This can been seen intuitively since the bare diffusion coefficient is the time integral of the Peculiar Velocity Autocorrelation Function of PVACF [5]. The PVACF is just the B-particle velocity with the total local fluid velocity subtracted and for large B-particles, the total local fluid velocity is equal to the B-particle velocity so the PVACF vanishes. The renormalization $R_\lambda^D[0]$ itself has been computed previously [18, 20] using mode coupling theory and a similar argument is used here to obtain $R_\lambda^D[0] = k_B T/(6\pi\eta[0] a/\xi)$ with the presence of $\xi$ to account for the transistion from slip to stick boundary conditions, the only new feature. The renormalized shear viscosity $\eta[0]$ appears in this result since the averaging length for the concentration is $\lambda > a$ for $\eta_\lambda \approx \eta[0]$ the usual renormalized shear viscosity. Combining this results together with $D_\lambda = 0$ in equation (12) yields the Millikan-Stokes-Einstein result

$$D[0] = \frac{k_B T}{6\pi\eta[0] a/\xi} \qquad (13)$$

This equation is in agreement with the Millikan formula [1] and it seems to agree with the Kim-Fedele data mean square displacement data as discussed in section 6. Again, equation (13) differs from Bedeaux-Mazur [15] and Keyes-Oppeneheim [20] just by the factor of $\xi$. Equation (13) can also be obtained using an Einstein-like argument [6] with the Fourier transform of equation (11) and this is instructive. Assuming a steady state with $d\bar{c}(r,t)/dt=0$ and $c(r,t) \propto \text{Exp}[-U(r)/(k_B T)]$ with U(r) the gravitational potential energy, one gets $D_\lambda = k_B T/(6\pi\eta_\lambda a/\xi)$. The limit $\lambda \to \infty$ of this expression together with the fact that $D_\lambda \to D[0]$ and $\eta_\lambda \to \eta[0]$ yields equation (13). $D_\lambda \to D[0]$ results form $\lambda \to \infty$ since in this limit, the local fluid velocity is zero and what remains is the ordinary velocity autocorrelation function and renormalized diffusion coefficient. This is not inconsistent with $D_\lambda = 0$ for large B-particles since for large B-particles the local fluid velocity is the same as the B-particle velocity and the PVACF vanishes causing $D_\lambda = 0$. The solution to equation (11) is easily obtained

$$\bar{c}_\kappa(t) = \text{Exp}\left[-D[\kappa]\kappa^2 t + i\kappa_z b_\lambda M g t\right] \qquad (14)$$

where $\bar{c}_\kappa(0) = 1$ which is the initial condition that the B-particle is at the origin and this will prove useful in sections 5 and 6.

## 5. The Fall Rate in Gravity

The average velocity of fall $U_{\text{fall}}$ or fall rate is obtained experimentally by measuring $\langle z \rangle_t$ which is the average B-particle displacement displacement $\langle z \rangle_t$ at time t is defined via

$$\langle z \rangle_t \equiv \int z \, c(r, t) \, dr \qquad (15)$$

$\hat{z}$ is in the direction of gravitational force Mg. Since $\bar{c}_\kappa(t)$ is the spatial Fourier transform of c(r, t) defined analogous to equation (2), it follows that

$$\frac{\partial}{\partial k_z}\bar{c}_\kappa(t) = i\int z \, c(r, t) \, e^{i\kappa \cdot r} \, dr \qquad (16)$$

and comparison of equations (15) and (16) yields the useful relation

$$\langle z \rangle_t = -i\left[\frac{\partial}{\partial \kappa_z}\bar{c}_\kappa(t)\right]_{\kappa=0} \qquad (17)$$

Expression (14) for $\bar{c}_\kappa[t]$ can easily be used in equation (17) to yield

$$\langle z \rangle_t = b_\lambda M g t \qquad (18)$$

Finally the fall rate $U_{\text{fall}}$ is given by taking the time derivative of equation (18)

$$U_{\text{fall}} = \frac{M g}{6\pi\eta_\lambda a/\xi} \qquad (19)$$

where equation (8) for the mobility $b_\lambda$ was utilized. Equation (19) is the same as the Millikan law of the fall rate *except* the bare shear viscosity $\eta_\lambda$ appears in equation (19) instead of the renormalized shear viscosity $\eta[0]$. The fall rate $U_{fall}$ given by equation (19) is graphed below together with the Kim-Fedele experimental data.

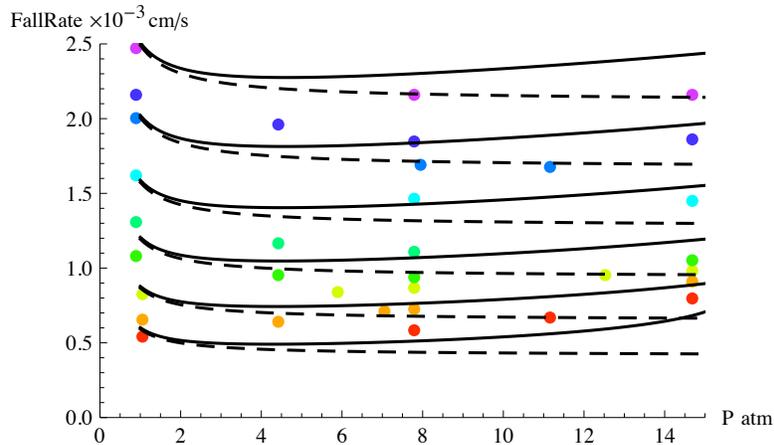

Figure: Experimental and theoretical fall rates as a function of gas pressure P

The solid lines in the graph above were obtained using our new fall rate formula equation (19) together with equation (5) for the bare shear viscosity and the averaging length was chosen as $\lambda=0.08 \times R$ to best fit the smallest radius a data. The solution to the quadratic equation (5) for the bare shear viscosity $\eta_\lambda$ has the virtue of fitting the smallest particles in the Kim-Fedele data better but the Pade approximant (6) allows for a much greater range of particle size and gas pressure. When equation (6) is used, the effect is smaller for the 0.45 $\mu M$ particles than shown in the figure above. The bottom solid line represents the smallest a=0.20 $\mu M$ particles and the top solid line represents the largest 0.45 $\mu M$ droplets with the intermediate solid lines in steps of 0.05 $\mu M$. The dots are the Kim-Fedele experimental data and dots of the same color represent droplets of the same size. The dashed lines are the result of the original Millikan fall rate formula which uses the usual experimental shear viscosity $\eta$. Clearly our new formula predicts the fall rate increases as the gas pressure is increases in qualitative agreement with the Kim-Fedele experimental results . This is in contrast to the Millikan result which predicts a decrease in the fall rate as the pressure is increased due to a transition from slip to stick boundary conditions. The new fall rate formula predicts a 45% enhancement over the Millikan fall rate for 0.20 $\mu M$ particles at 15 atm. The smaller the B-particles the greater the effect of the fluctuation renormalization on the shear viscosity because the fluctuations are larger in comparison with the particle. The effect of the bare viscosity is nonlinearly dependent on the particle radius and the graph shows the effect of the bare shear viscosity appears rather "explosively" as the radius of the particles studied is decreased.

There is another way to analyze the experimental fall rate data. The fall rate can be used together with equation (19) to determine the bare shear viscosity so the apparatus functions as a sort of bare shear viscometer. This $\eta_\lambda$ can then be used with, for exmaple, equation (6) to solve for $\lambda$. This value of $\lambda$ together with the bare shear viscosity $\eta_\lambda$ could then be used in a different experiment like, for example, the motion of a sub-micron size particle in a time varying or constant electric field and or magnetic field.

# 6. The Mean Square Displacement

The deviation from the average distance of fall is defined as $\delta z = z - \langle z \rangle$ and then the average of the square of the deviation is $\langle \delta z^2 \rangle_t = \langle z^2 \rangle_t - \langle z \rangle_t^2$ where $\langle z \rangle_t$ has already been calculated in equation (18). $\langle z^2 \rangle_t$ is defined in a manner similar to $\langle z \rangle_t$ in equation (15) and $\langle z^2 \rangle_t$ is easily obtained by taking a second derivative with respect to the wavenumber component $\kappa_z$

$$\langle z^2 \rangle_t = -\left[\frac{\partial^2}{\partial \kappa_z^2} \overline{c}_\kappa(t)\right]_{\kappa=0} \tag{20}$$

Using $\overline{c}_\kappa(t)$ from equation (14) in equation (20) together with equation (18) yields the familiar Einstein formula

$$\langle \delta z^2 \rangle = 2 D[0] t \tag{21}$$

where the renormalized diffusion coefficient D[0] is given by equation (13) in terms of $\eta[0]$ the bare shear viscosity. $\langle \delta z^2 \rangle_t$ was measured by Kim-Fedele while doing the fall rate measurements and they essentially showed their data for $\langle \delta z^2 \rangle_t$ was in agreement with equation (21) combined with equation (13) for D[0]. Thus it possible to explain both the abnormal fall rates $U_{fall}$ and the normal behavior of $\langle \delta z^2 \rangle_t$ with the same theory. The basic reason for this is that the current $\mathbb{J}$ in equation (10) depends upon the bare diffusion coefficient $D_\lambda$ and the bare mobility $b_\lambda$. The subsequent "fluctuation-renormalization" of the convective term in equation (9) results in the bare diffusion coefficient being renormalized D[0] in equation (11) while the bare mobility is not renormalized.

## 7. Summary and Conclusions

The theory of the bare shear viscosity $\eta_\lambda$ described here was used above to explain the Kim-Fedele fall rate data and the results are in reasonable agreement given what is know experimentally and theoretical at the present time. The fall distance measurements $\langle z \rangle_t$ were taken at 10 seconds in the Kim-Fedele experiments and for the smallest particles studied, and the $\langle z \rangle_t$ was comparable in with $\sqrt{\langle \delta z^2 \rangle}$. Taking the fall distance measurements over a longer time would help to reduce the experimental uncertainty in the fall rates but at a cost in terms of the time the experiment would run. Fuchs [1, 21] and others have pointed out the special problems of measuring the fall rate for particles so small that their Brownian motion is large. The utility of the bare shear viscosity $\eta_\lambda$ concept will become more apparent as it is used for prediction in future experiments like, for example, the fall rate of charged submicon droplets in an electric or magnetic field. Also the fall rate of smaller sub-micron sized particles could be measured at pressures higher than 15 atm. Fall rates of particles under pressure greater than one atm would be important in the study of the atmosphere of planets like Saturn and Jupiter. Also a Langevin equation based argument can be used to provide an explanation of the Kim-Fedele experiment and but this will be presented elsewhere.

## Acknowledgement

It is a pleasure to thank Yong Kim and Paul Fedele for many discussions experiments discussed here. Their description of the details of the experiment made a great impression on me especially the care with which the experiments were performed as well as some of the difficulties that were overcome. The author also thanks the Faculty Fellowship Leave Program of the Hunter College of the City University of New York during the academic year 1981-1982 which allowed the author to spend a good deal of time at Lehigh University.

## References and Notes

1. *The Mechanics of Aerosols*, by N.A. Fuchs (Dover, NY, 1989) org. pub. Pergamon Press (1964).
2. L.W. Alvarez, W. Alvarez, F. Asaro, and H.V. Michel, Science 208, 1095 (1980).
3. R. Millikan, Phys. Rev. 22, 1 (1923).
4. Y.W. Kim and P.D. Fedele, Phys. Rev. Lett. 48, 403 (1982).
5. R.L. Varley, Fluc. Noise Lett. 6, L179 (2006); R.L. Varley, arXiv:cond-mat/0402422v3 [cond-math.stat-mech]; R.L. Varley and G. Sandri, Physics Letters A 124, 411 (1987).
6. *Fluid Mechanics*, L.D. Landau and E.M. Lifshitz (Addison-Wesley, Reading, Mass., 1959) esp. sec. 1, 20, 58, 59, and 133.
7. D.N. Zubarev and V.G. Morozov, Physica 120A, 411 (1983) esp. 413-414.
8. *The Mathematica Theory of Nonuniform Gases*, S. Chapman and T.G. Cowling (Cambridge U Press, 1970). esp. p. 26-27.
9. *Theoretical Microfluidics*, H. Bruus (Oxford U. Press, 2008).
10. *Nonequilibrium Statistical Mechanics*, R. Zwanzig, (Oxford U. Press, 2001).
11. T.Keyes and I. Oppenheim, Phys. Rev. 7A, 1384 (1973).
12. K. Kawasaki, Ann. Phys. (NY) 61, 1 (1970).
13. H. Mori and H. Fujisaka, Prog. Theo. Phys. (Japan) 49, 764 (1973) esp. (2-17), (2-19) and (2-21) or (1-11).
14. R. Zwanzig, Nonlinear Dynamics of Collective Modes, in Proc. 6th IUPAP Conf. on Stat. Mech., S. Rice et. al. (eds.) (U Chicago, 1972); I. Oppenheim, Transport Theory and Stat. Phys. 24, 781 (1995); T.R. Kirkpatrick and J.C. Nieuwoudt, Phys. Rev. 33A, 2651 (1986); J. Böse, W. Götze, and M Lücke, Phys. Rev. 17, 434 (1978).

15. P. Mazur and D. Bedeaux, Physica 76, 235 (1974); P. Mazur and G. van der Zwan, Physica 92A, 483 (1978); 98, 169 (1979); R.L. Varley and R.L. Zhou, Physica A 127 (1984) 363.
16. P. Epstein, Phys. Rev. 23, 710 (1924).
17. M.P. Allen and O.G. Raabe, J. Aersol Sci 13, 537 (1982).
18. D. Bedeaux and P. Mazur, Physica 73, 431 (1974).
19. *Non-Equilibrium Thermodynamics,* S.R. de Groot and P. Mazur (Dover, NY, 1984) org. pub. North-Holland, Amsterdam, 1962 esp. chap. XI.
20. T. Keyes and I. Oppenheim, Phys. Rev. 8A, 937 (1973).
21. *Stochastic Proceses in Physics and Chemistry*, N.G. van Kampen (North-Holland, Amsterdam, 1981).
22. *Multiple time scale calculations of the velocity autocorrelation function: Hard-disk results*, Xing Hong (CUNY Graduate Center, MA thesis, 2008) unpublished.